\documentclass[conference]{IEEEtran}
\usepackage{graphicx,graphics}
\usepackage{wasysym}
\usepackage{csquotes}
\usepackage{cases}
\usepackage{cite}
\usepackage{array}

\begin{document}

\title{Synchronization Detection in Networks of Coupled Oscillators for Pattern Recognition}

\author{\IEEEauthorblockN{
		Damir Vodenicarevic\IEEEauthorrefmark{1}
		Nicolas Locatelli\IEEEauthorrefmark{1},
		Julie Grollier\IEEEauthorrefmark{2}, 
		and Damien Querlioz\IEEEauthorrefmark{1}}
	
	\IEEEauthorblockA{\IEEEauthorrefmark{1}Institut d'Electronique Fondamentale, Univ. Paris-Sud, CNRS, 
		91405 Orsay, France\\ 
		Email: damir.vodenicarevic@u-psud.fr, damien.querlioz@u-psud.fr}
	\IEEEauthorblockA{\IEEEauthorrefmark{2}Unite Mixte de Physique CNRS/Thales and Universite Paris-Sud, 1 Ave. A. Fresnel, 91767 Palaiseau, France}
}

\maketitle

\begin{abstract}
Coupled oscillator-based networks are an attractive approach for implementing hardware neural networks based on emerging nanotechnologies.
However, the readout of the state of a coupled oscillator network is a difficult challenge in hardware implementations, as it necessitates complex signal processing to evaluate the degree of synchronization between oscillators, possibly more complicated than the coupled oscillator network itself.
In this work, we focus on a coupled oscillator network particularly adapted to emerging technologies, and evaluate two schemes for reading synchronization patterns that can be readily implemented with basic CMOS circuits.
Through simulation of a simple generic coupled oscillator network, we compare the operation of these readout techniques with a previously proposed full statistics evaluation scheme.
Our approaches provide results nearly identical to the mathematical method, but also show better resilience to moderate noise, which is a major concern for hardware implementations.
These results open the door to widespread realization of hardware coupled oscillator-based neural systems.
\end{abstract}


\bstctlcite{IEEEexample:BSTcontrol}

\section{Introduction}

The interest for neuro-inspired computing architectures is currently rising considerably, due to the inefficiency of the von Neumann architecture for the treatment of cognitive tasks like recognition or classification of massive amounts of data~\cite{yen-kuang_chen_convergence_2008, shibata_bio_2009, indiveri_memory_2015, querlioz_bioinspired_2015}.
Both academic and industrial groups are focusing efforts on developing efficient physical implementations of artificial neural network architectures, emulating neuronal and synaptic behavior by the means of complex CMOS circuits \cite{merolla_million_2014, benjamin_neurogrid_2014}, or relying on the physical properties of emergent technologies, such as spintronics \cite{roy_brain-inspired_2014, vincent_synapse_2015}, oxide devices \cite{pershin_experimental_2010,  prezioso_training_2014, saighi_plasticity_2015} or phase change devices \cite{bichler_visual_2012}.

Inspired by the importance of temporal patterns of activity in neural assemblies in the brain achieving cognitive functions~\cite{axmacher_memory_2006, bhowmik_how_2012}, an alternative cognitive architecture consists of building networks of coupled oscillators.
The rich dynamics of coupled oscillators can be leveraged to operate classification and recognition tasks~\cite{hopfield_neural_1982, johannet_specification_1992, hoppensteadt_associative_1997, izhikevich_weakly_1999, nikonov_coupled-oscillator_2015, levitan_associative_2013}, based on the emergence of synchronization patterns among the oscillators of the network. 
Beyond theoretical studies, experimental implementations of cognitive networks of oscillators have been proposed based on numerous technologies such as CMOS differential oscillators~\cite{cosp_synchronization_2004}, CMOS ring oscillators~\cite{sivilotti_novel_1990, shibata_cmos_2012, cotter_computational_2014}, or laser oscillators~\cite{hoppensteadt_synchronization_2000}.
Additionally, emerging nanotechnologies provide nanometer-scaled oscillators that are exceptionally compact, have easily tunable frequencies, can be very fast, and show synchronization capabilities that may be used for oscillator-based computing. Examples include spin-torque nano-oscillators~\cite{csaba_spin_2012, nikonov_coupled-oscillator_2015, fan_injection_2015} or oxide-based relaxation oscillators~\cite{datta_neuro_2014, shukla_pairwise_2014, parihar_exploiting_2014, sharma_phase_2015}. 

Most of these proposals are based on Hopfield’s model, where the construction of the cognitive capabilities of the network (learning) happens by tuning the connections between each pair of oscillators~\cite{hopfield_neural_1982, johannet_specification_1992}. With emerging nanotechnologies, controlling oscillator coupling with high accuracy can be a difficult challenge. As a consequence, other schemes for computing with coupled oscillators, where the network has fixed connections that do not need to be controlled perfectly, and oscillators have adjustable natural frequencies, are especially attractive as they would face less difficulties with hardware
implementation~\cite{vassilieva_learning_2011, shibata_cmos_2012, nikonov_coupled-oscillator_2015}, and map well to emerging technologies~\cite{csaba_spin_2012, datta_neuro_2014}.


In particular, in~\cite{vassilieva_learning_2011}, Vassilieva \textit{et al.} propose a weakly coupled oscillator network for pattern recognition where the coupling strengths between oscillators do not need to be tuned, and the cognitive operations are performed by only changing the natural frequencies of the oscillators.
This network is also relatively resilent to oscillator phase noise, a major concern of oscillators based on nanotechnologies.
Unfortunately, the readout of such a network remains a challenge from a hardware implementation perspective.
It involves the detection of synchronization patterns, and requires the evaluation of synchronization between each pair of oscillators.
Achieving this readout with fast, possibly noisy oscillators in real time might even be more complicated than the pattern recognition itself.

Therefore, in this study, we propose and investigate simple, easy to implement schemes for the evaluation of the degree of synchronization between pairs of oscillators, in the context of the oscillator-based recognition architecture of~\cite{vassilieva_learning_2011}.
First, we describe the weakly coupled oscillator network architecture following the proposal of~\cite{vassilieva_learning_2011} and introduce three protocols for the evaluation of the degree of synchronization between each pair of oscillators.
Then, we present and compare the recognition capabilities of the oscillator network obtained when using each readout protocol, in the case of a network of noiseless oscillators as well as in a more realistic case of noisy oscillators. 
We conclude on the potential for large scale integration of each protocol.

\section{Description of the Considered System}
\label{sec:system}
\subsection{An Oscillator Network for Pattern Recognition}
\label{subsec:descr_network}

\begin{figure}[h]
	\centering
	\includegraphics[width=\linewidth]{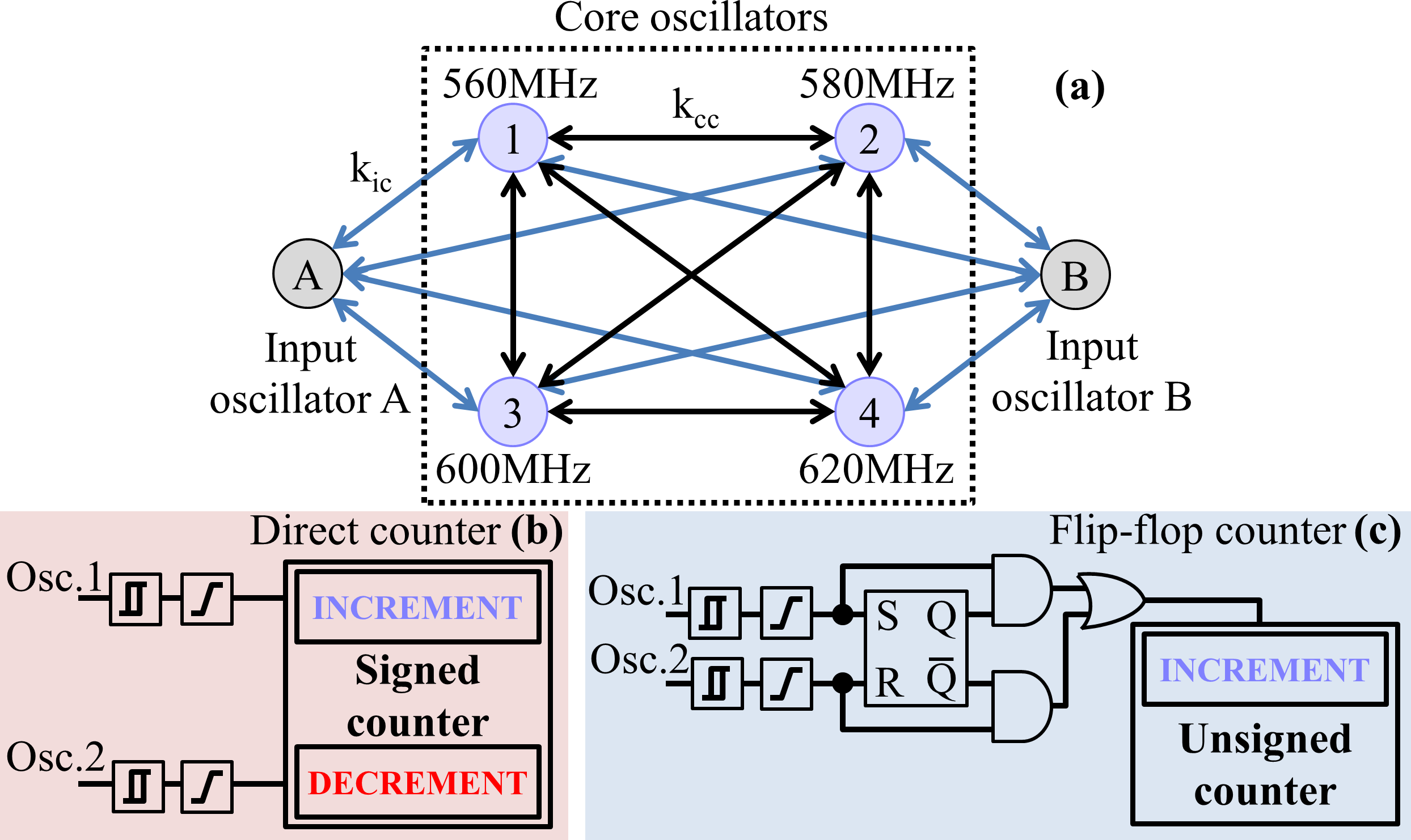}
	\caption{(a) Illustration of an oscillator network architecture for pattern recognition operation, showing core oscillators, input oscillators, and the corresponding couplings. (b,c) Illustration of proposed schemes for evaluating the degree of synchronization between pairs of oscillators: (b) direct counter evaluation technique, (c) flip-flop counter evaluation technique.}
	\label{fig:architecture}
\end{figure}

As proposed in~\cite{vassilieva_learning_2011}, we use a network of weakly coupled oscillators for pattern recognition. The considered network is based on a set of oscillators whose natural frequencies are controlled and set by external means (voltage, current, etc.). For the purpose of our demonstration, and as illustrated in Fig.~\ref{fig:architecture}(a), the core of the architecture is a small network of four coupled oscillators labeled \{1,2,3,4\}. This network is connected to two input oscillators \{A,B\}.
Such an architecture is able to classify analog patterns $\{s_{A},s_{B}\}$ coded as input oscillators' natural frequencies $\{f_0^{(A)},f_0^{(B)}\}$ by associating them to different synchronization patterns within the core network.
Indeed, the combined influence of the input oscillators induces computation among the core oscillators, by bringing their complex dynamics to converge to given synchronization states.
The coupling strengths, together with the natural frequencies of the oscillators, can be tuned by learning algorithms in order to shape the response of the network and classify a desired set of patterns~\cite{vassilieva_learning_2011}.

All the oscillators are modeled using the Kuramoto equation:
\begin{equation}
	\dot{\varphi_{n}} = 2 \pi f_0^{(n)} + 2 \pi \sum\limits_{m \neq n} k_{mn} \sin(\varphi_{m}-\varphi_{n}) + \eta
	\label{eq:kuramoto}
\end{equation}
where $\varphi_{n}$ is the phase of oscillator $n$, $f_0^{(n)}$ is its natural frequency and the $k_{mn}$ matrix defines the coupling between the oscillators of the network. $\eta$ is a Gaussian random noise term with a standard deviation defined by $\sqrt{2\pi \cdot \mathrm{FWHM}/dt}$ so that isolated oscillators' power spectra have a Full-Width at Half Maximum FWHM.
These stochastic differential equations were simulated using the Milstein method \cite{milshtejn_approximate_1975} with a timestep of $100ps$.

To build the recognition network, each core oscillator is coupled to all the others, while input oscillators are coupled to core oscillators only. In this study, we focus on a simple case where the input-core coupling strengths and the core-core coupling strengths are uniform and defined by $k_{ic}$ and $k_{cc}$ respectively. All couplings are considered bidirectional ($k_{nm}=k_{mn}$).
The natural frequencies of the core oscillators are assumed to be equally distributed. Unless otherwise stated, we assume that $k_{ic} = 12$\,MHz, $k_{cc} = 4\,MHz$ and the natural frequencies of the core oscillators are set to $\{560, 580, 600, 620\}$\,MHz.

\begin{figure*}[ht]
	\centering
	\includegraphics[]{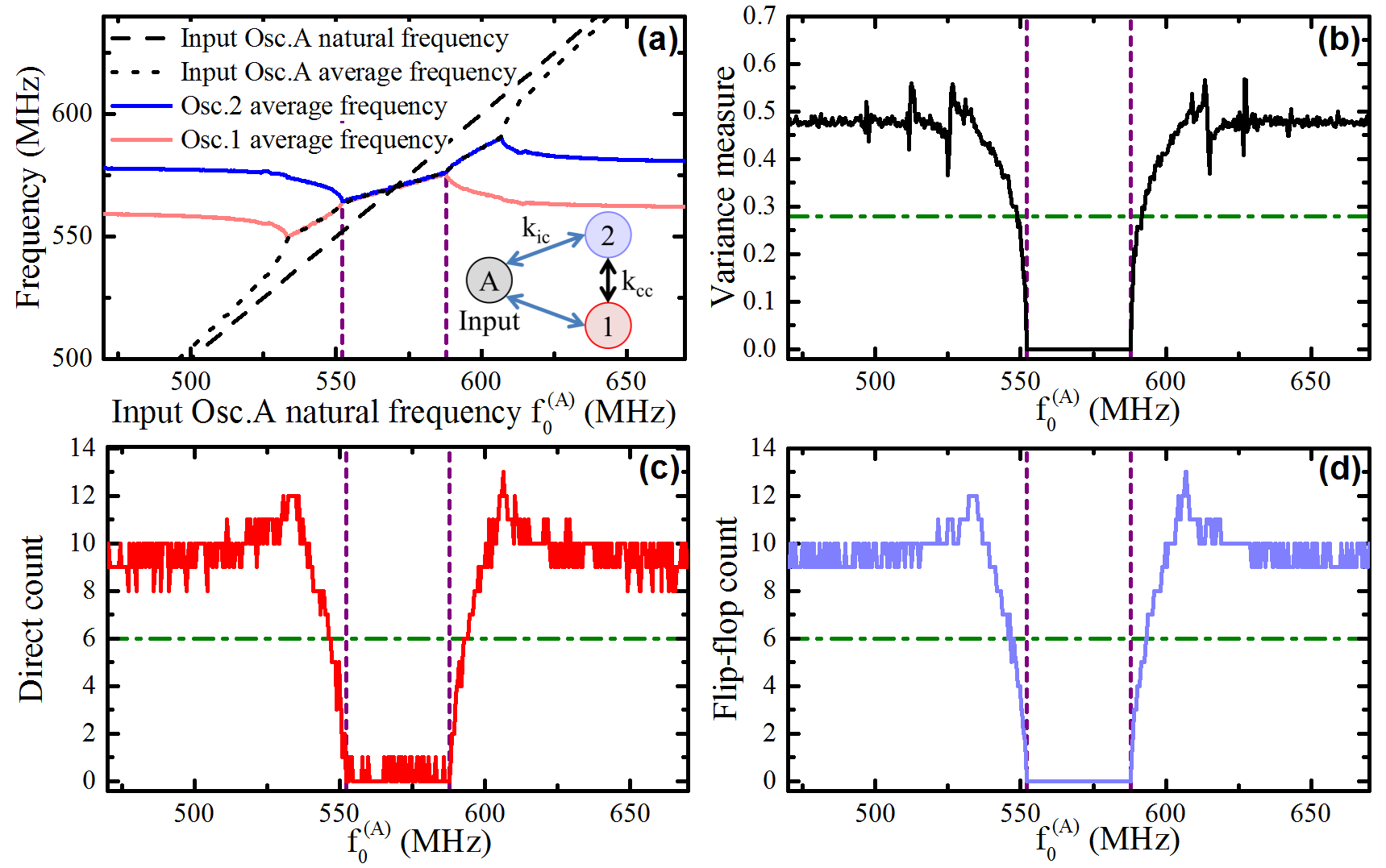}
	\caption{Simulations in a simplified, noiseless situation with two core oscillators with fixed natural frequencies $\{f_{0}^{(1)},f_{0}^{(2)}\}=\{560,580\}$\,MHz, and one input oscillator whose natural frequency $f_0^{(A)}$ is varied.  (a) Average frequencies of the three oscillators as a function of $f_0^{(A)}$. The quasi-synchronization of the two core oscillators is evaluated using the three readout schemes, and their outputs are plotted as a function of $f_0^{(A)}$ for (b) variance measure, (c) direct counter, and (d) flip-flop counter schemes. Vertical purple dotted lines indicate the range of perfect synchronization between oscillators 1 and 2. Horizontal green dotted lines correspond to selected thresholds under which oscillators are declared quasi-synchronized.}
	\label{fig:SchemesOutputs}
\end{figure*}

\subsection{Schemes for Evaluating Synchronization Patterns}

The readout process of such an architecture involves pairwise synchronization detection between core oscillators, and the output is the resulting list of synchronized pairs. In weak coupling conditions and especially in the presence of phase noise, synchronization between oscillators in its strictest definition is hardly achieved, as perfect phase-locking seldom occurs. Thus, a weaker definition of synchronization is needed and should be based on a measure of the degree of synchronization between two signals.
For instance, statistical methods such as the variance measure introduced in~\cite{vassilieva_learning_2011} can be used to define quasi-synchronization between a pair of oscillators \{n,m\}:
\begin{equation}
\mathrm{Var}_\tau(\sin(\varphi_n-\varphi_m)) < \epsilon_v
\label{eq:eq_var}
\end{equation}
where the variance is evaluated during a limited time $\tau$. A threshold $\epsilon_v\in[0;0.5] $ is chosen to discriminate quasi-synchronized pairs from non-synchronized pairs, $\epsilon_v=0$ corresponding to a perfect synchronization requirement.
This definition was shown to allow the detection of a rich set of synchronization patterns in weakly coupled networks \cite{vassilieva_learning_2011}.
However, the hardware implementation of such a detection scheme would involve a combination of complex circuits, or an external computing unit: it is not a reasonable readout technique for an efficient physical implementation.

Here, we propose two other quasi-synchronization detection schemes based on operational principles compatible with an easy CMOS implementation and compare their performances with the variance-based method. The oscillator signals are digitized using Schmitt triggers, intended to increase the noise resilience of the readout, after which a rising edge detection is performed to obtain a single pulse per period of the signals. 

A first synchronization evaluation approach that is further called ``direct counter'' is presented in Fig.~\ref{fig:architecture}(b). This technique aims at evaluating the difference $\Delta N_{\tau}$ between the number of periods of the two signals during a given amount of time $\tau$. The counting is achieved by incrementing or decrementing a counter at each rising edge of the respective signals. The result is then compared to a threshold $\epsilon_d$, and the two oscillators are considered synchronized if $|\Delta N_{\tau}|<\epsilon_d$. 

A second scheme that is further called ``flip-flop counter'' is presented in Fig.~\ref{fig:architecture}(c). This technique exploits the fact that if two signals are synchronized, their rising edges should alternate. A counter is then incremented each time two consecutive rising edges of the same signal are not separated by a rising edge of the second. Again, after the evaluation time $\tau$, the two oscillators are considered synchronized if the final value of the counter is strictly lower than a given threshold $\epsilon_f$.

Because they use only limited information from the signals, the two counter-based schemes differ conceptually from the variance measure which depends on the full time dependence of the phase difference between the two oscillators. 
We also highlight the major difference between the two counter-based schemes: while the direct counter method measures an average frequency difference during the total evaluation time, the flip-flop counter takes into account and sums up every detected local desynchronization event. Yet, the physical implementation of the flip-flop counter would require less components, mainly because of the simpler, unsigned counter it uses.

\subsection{Equivalence of the Detection Schemes}
\label{sec:comparison_perfs}

To compare the synchronization evaluation schemes, we first investigate the simplified case of a single input oscillator $\{A\}$ and two core oscillators $\{1,2\}$, as illustrated in Fig.~\ref{fig:SchemesOutputs}(a). Core oscillators' natural frequencies are set to $\{f_{0}^{(1)},f_{0}^{(2)}\}=\{560,580\}$\,MHz and the input oscillator's natural frequency $f_0^{(A)}$ is swept from $470$ to $670$\,MHz. 
Fig.~\ref{fig:SchemesOutputs}(a) captures the synchronization phenomenon between the three oscillators by showing the evolution of their average frequencies.
While the coupling between core oscillators \{1,2\} is initially too weak for them to synchronize, they are eventually brought to synchronization when the input oscillator's natural frequency lies in a limited range.

For every simulation, the oscillator network dynamics are computed for $1\mu s$. After a $0.5\mu s$ cool-down time to wait for the convergence of the network dynamics, the three detection schemes are evaluated during $\tau=0.5\mu s$ between core oscillators \{1,2\}. Their outputs are plotted in Figs.~\ref{fig:SchemesOutputs}(b,c,d) before the thresholding operation.

The three curves appear extremely similar. It is surprising to note that, while variance measure and counter approaches use different basic principles, the obtained curves can actually almost be superimposed.
All exhibit a very distinct dip to a zero-value when the two core oscillators are synchronized, and a high plateau value when the oscillators are desynchronized.
In the intermediate range, where the oscillators are quasi-synchronized, the outputs show a progressive increase, allowing to define thresholds that will discriminate whether the oscillators are quasi-synchronized or not.

From these curves, we choose equivalent thresholds for the three detection schemes: $\epsilon_{v}=0.28$ for the variance measure scheme, $\epsilon_{d}=6$ for the direct counter scheme, and $\epsilon_{f}=6$ for the flip-flop counter scheme. In the following  we use these threshold values if not stated otherwise.


\begin{figure*}[ht]
	\centering
	\includegraphics[]{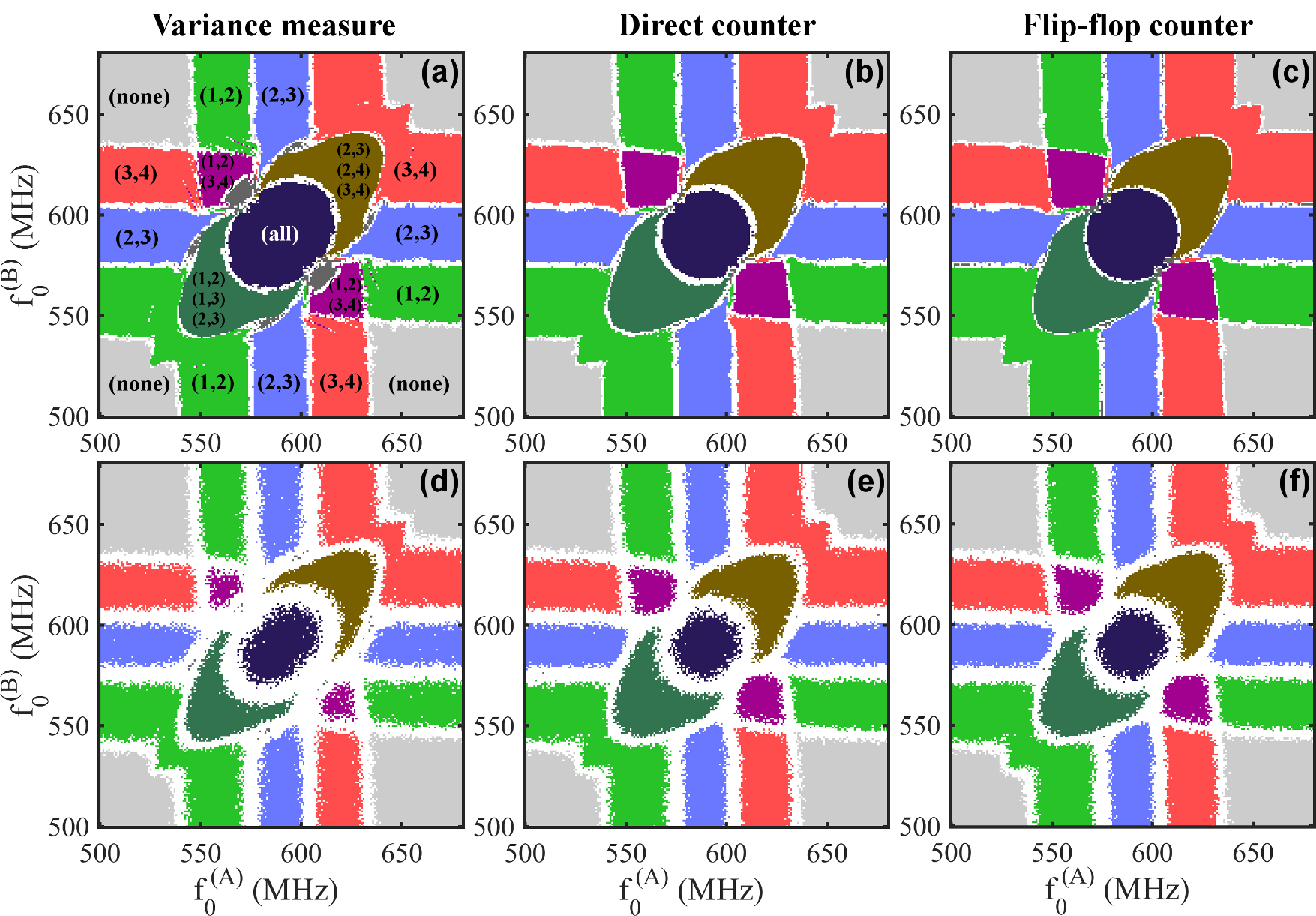}
	\caption{Readout maps showing the distribution of synchronization patterns among core oscillators as a function of the two input oscillators' natural frequencies ${f_{0}^{(A)},f_{0}^{(B)}}$, as detected by the three readout protocols.  Each color is associated to a single synchronization pattern, as specified on Fig.(a). The (a,b,c) maps are obtained in a situation with noiseless oscillators, (d,e,f) maps are obtained for oscillators with phase noise corresponding to FWHM=1\,MHz. Readout maps are evaluated respectively using: (a,d) the variance measure scheme, (b,e) direct counter scheme, and (c,f) flip-flop counter scheme.}
	\label{fig:patterns}
\end{figure*}

\section{Pattern Recognition and Comparison of the Readout Schemes}
\label{sec:results}

We now evaluate the  three readout schemes on the full coupled oscillator network of Fig.~\ref{fig:architecture}(a), introduced in section~\ref{subsec:descr_network}. 

\subsection{Readout Maps in the Absence of Noise}
\label{sec:equivalence}

Figs.~\ref{fig:patterns}(a,b,c) present the readout maps of the synchronization patterns in the core network, as a function of the input oscillators' natural frequencies $\{f_0^{(A)},f_0^{(B)}\}$ in the case of noiseless oscillators ($\eta=0$). They are obtained using the variance measure, direct counter and flip-flop counter detection schemes respectively.

For each point in this map, the oscillator network dynamics are simulated for $1\mu s$.
After a $0.5\mu s$ cool-down time to wait for the convergence of the network dynamics, the three detection schemes are performed on each of the six pairs of core oscillators during $\tau=0.5\mu s$, and the results are compared to their respective thresholds. For each simulation, an output list of synchronized pairs is then given by each readout scheme.
To account for the robustness of the readout results to initial conditions, each point on the map is simulated ten times with random initial phases. 
If the ten simulations do not result in the same output synchronization pattern, the point is discarded as ``inconsistent'' and left blank on the map. If the ten simulations yield identical results, the point is then colored on the map according to the output pattern.

Producing these $200\times 200$-point maps, to allow precise assessment of the coupled oscillator network behavior, comes at a high computational cost as it requires $200\times 200\times 10=400,000$ independent simulations per map. For optimal efficiency, the simulations were performed on a nVidia Tesla K20 GPU, using the Cuda Thrust C++ library.

In this noiseless example, the three evaluation schemes yield rich output maps, with large and well-defined regions associated to different synchronization patterns. The boundary regions (blank) where no repeatable readout is obtained are relatively small.
These results show the efficient recognition capability of the oscillator network, as already pointed out in~\cite{vassilieva_learning_2011}. Indeed, it spontaneously discriminates  inputs through the establishment of synchronization patterns in its core. Additionally, it is remarkable that  all of the introduced synchronization detection schemes are operational and lead to highly similar readout maps. 

\begin{figure}[h]
	\centering
	\includegraphics[]{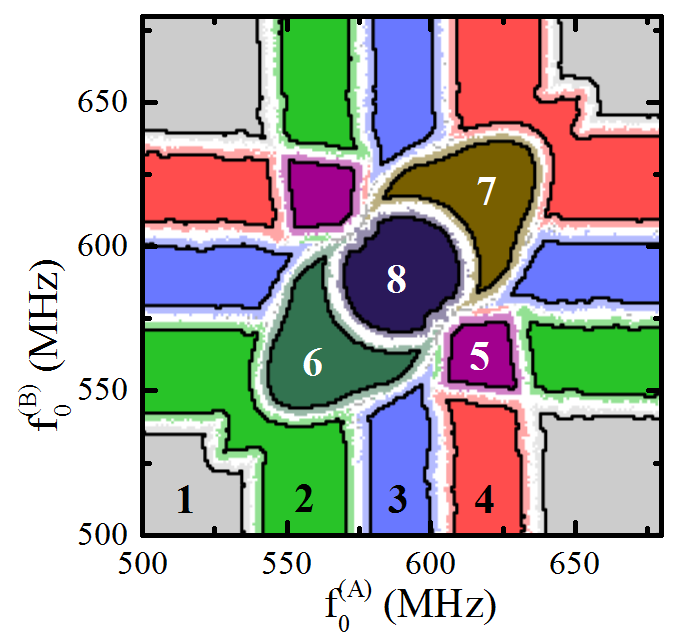}
	\caption{Effect of the filter applied to the readout maps for counting discriminated patterns. This is illustrated in the case of the readout map obtained for a noiseless network and the flip-flop counter detection scheme (see Fig.~\ref{fig:patterns}(c)). Saturated regions delimited by a black line are considered robust and kept by the filter, pale areas are ignored. The numbers on the map index the eight unique discriminated patterns.}
	\label{fig:Filtre_Boule}
\end{figure}

The capabilities of the oscillator network associated with each readout scheme are evaluated through the number of classes of patterns the architecture is able to discriminate, \textit{i.e.} the number of regions with different readouts that appear on the map.
In this counting, we choose to ignore isolated points, as well as porous regions where consecutive points do not consistently yield the same output. To do so, a filter is applied on the readout maps, as illustrated on Fig.~\ref{fig:Filtre_Boule}, that only keeps regions which yield identical outputs in an at least $3$\,MHz radius area. This ensures that the counted classes are represented by large and continuous regions that are tolerant to small input variations.

In the case of both counter-based readouts, eight patterns are discriminated, each one being associated to a different synchronization pattern. Meanwhile, a ninth output pattern appears for the variance-based readout.

\subsection{Readout Maps in the Presence of Noise}

When considering hardware implementations, especially based on nanotechnology such as proposed in
\cite{csaba_spin_2012, nikonov_coupled-oscillator_2015, fan_injection_2015,datta_neuro_2014, shukla_pairwise_2014, sharma_phase_2015}, the consequences of the phase noise of the oscillators, which can significantly perturb the network dynamics, needs to be considered.
To account for it, we reproduced the simulations of the network dynamics as well as the three readout schemes including a non zero noise corresponding to oscillators' $\mathrm{FWHM}=1$\,MHz.

We show in Fig.~\ref{fig:patterns}(d,e,f) the readout maps obtained for the noisy oscillator network.
Again, for all three detection schemes, the three maps remain very similar. Compared to the noiseless network, the class regions are sensibly reduced, and the blank (inconsistent) regions are getting wider, as could be expected.
Indeed, as the noise increases, the repeatability of the readouts becomes an increasing issue.
It notably has an impact on the readout map obtained through the variance-based scheme. As one can see in Fig.~\ref{fig:patterns}(d), the ninth synchronization pattern is no longer observed, and the purple region has become particularly porous.

\subsection{Testing Detection Schemes Against Different Network Parameters}

\begin{figure}[h]
	\centering
	\includegraphics[]{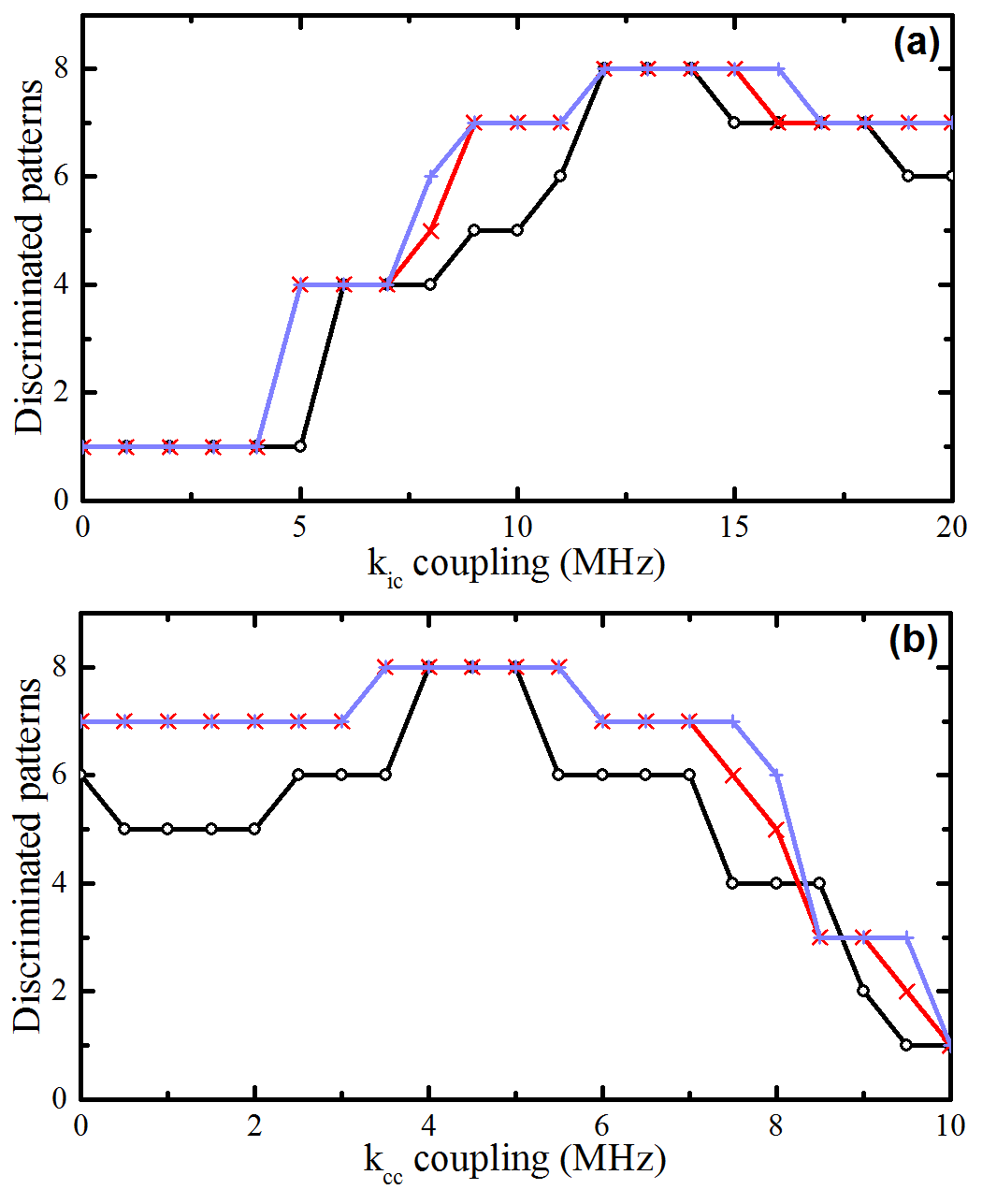}
	\caption{Number of discriminated patterns within the core network as a function of (a) input-core $k_{ic}$ coupling and (b) core-core $k_{cc}$ coupling as evaluated using the three detection schemes. Simulations performed with $\mathrm{FWHM}=1$\,MHz.}
	\label{fig:Patterns_Vs_Couplings}
\end{figure}

Modifying the coupling strengths between oscillators in the network  changes the distribution of synchronization patterns, and also has an influence on the relative phase dynamics between coupled oscillators, and subsequently on the synchronization readout.
To compare  the detection schemes in the case of other network configurations, we repeated the simulations of noisy networks with varying input-core coupling strengths $k_{ic}$ and core-core coupling strengths $k_{cc}$. Fig.~\ref{fig:Patterns_Vs_Couplings}(a) shows the number of discriminated classes of inputs when $k_{cc}=4$\,MHz is kept constant and $k_{ic}$ is varied. Fig.~\ref{fig:Patterns_Vs_Couplings}(b) shows the number of discriminated classes of inputs when $k_{ic}=12$\,MHz is kept constant and $k_{cc}$ is varied.

The two counter-based readout schemes provide quasi-identical results. Both plots show that optimal coupling strengths, for which a maximum number of classes can be discriminated, fall in the same range for the three readout schemes. However, the counter-based schemes lead to the discrimination of a higher number of classes than the variance-based one in large ranges of coupling strengths.
These observations suggest that the counter-based definitions of quasi-synchronization might be more robust than the variance-based one. 
We discuss and interpret this idea in the next section.

\section{Sensitivity of the Readout Schemes to Noise and Parameters}
\label{sec:comparison}

In this section, we perform an in-depth evaluation and comparison of these schemes, and discuss their applicability for a final hardware system.

\subsection{Resilience to Noise of the Readout Schemes}
\label{sec:ResToNoise}

\begin{figure}[h]
	\centering
	\includegraphics[]{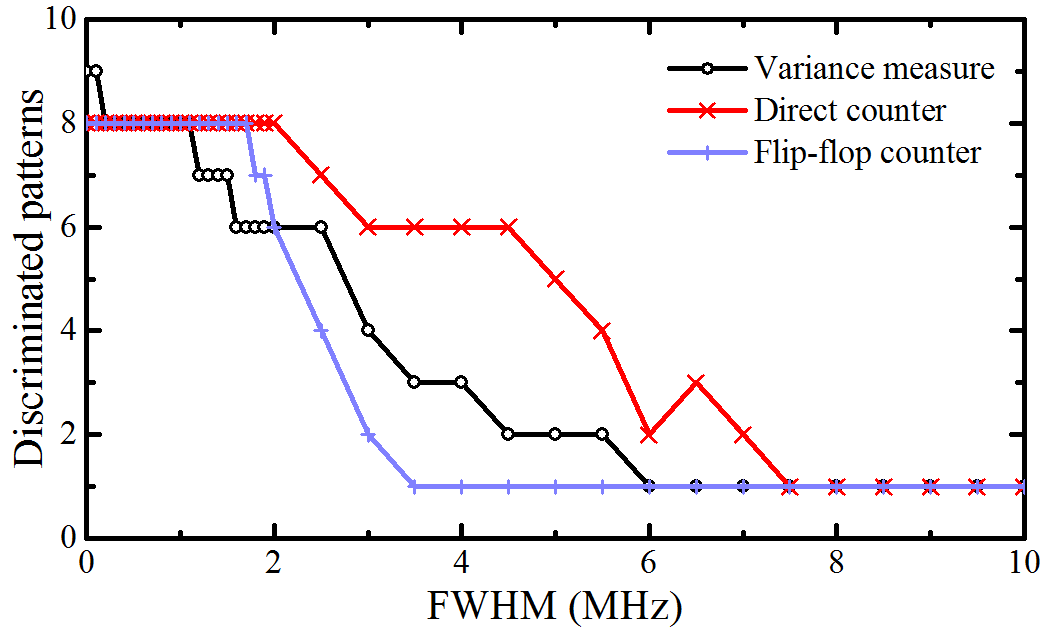}
	\caption{Number of discriminated patterns as a function of oscillators' FWHM, as evaluated using the three detection schemes.}
	\label{fig:Patterns_Vs_Noise}
\end{figure}

To assess the relative influence of phase noise on the three readout schemes, we repeat the simulations of the noisy network with increasing noise levels and plot the evolution of the number of classes discriminated by each readout method in Fig.~\ref{fig:Patterns_Vs_Noise}. The three schemes show different resiliences to noise. 

For low noise levels, the variance-based scheme shows the lowest resilience, as it is the first detection scheme that stops being able to discriminate eight classes. Indeed, the variance measure is strongly affected by fluctuations appearing in the phase difference dynamics. Moreover, an identical number of phase-slip events can have dramatically different consequences depending on the synchronization recovery duration. Therefore, when looking at the outputs before thresholding, we observe an increasing spreading of the results for the ten repeated simulations with increasing phase noise. This induces spurious detections of synchronization or desynchronization producing many inconsistent points, which leads to the disappearance of some classes in the readout map.

The two counter-based schemes keep their ability to discriminate eight patterns for higher noise levels, even beyond $2$\,MHz for the direct counter scheme. They evaluate the exact number of desynchronization events and are not sensitive to their dynamics.
However, the flip-flop counter scheme appears to fail very rapidly when the FWHM of the oscillators goes above $2$\,MHz, even faster than the variance-based scheme. In the flip-flop counter detection method, every desynchronization event, \textit{i.e.} phase slip between the two signals, is detected and counted. As the noise level increases, many spurious desynchronization events get detected, which eventually leads the counter to go above the threshold. Only strong synchronizations are then detected. This observation suggests that the threshold level should be raised to adapt to high phase noise oscillator networks.

Because it simply evaluates an average frequency difference between the two noisy signals, the direct counter scheme is the one showing the best resilience to noise. It is still able to discriminate six classes of inputs for oscillators with FWHMs up to 4.5\,MHz, when other readout methods only discriminate one or two classes.


\subsection{Influence of the Choice of the Threshold }

\begin{figure}[h]
	\centering
	\includegraphics[]{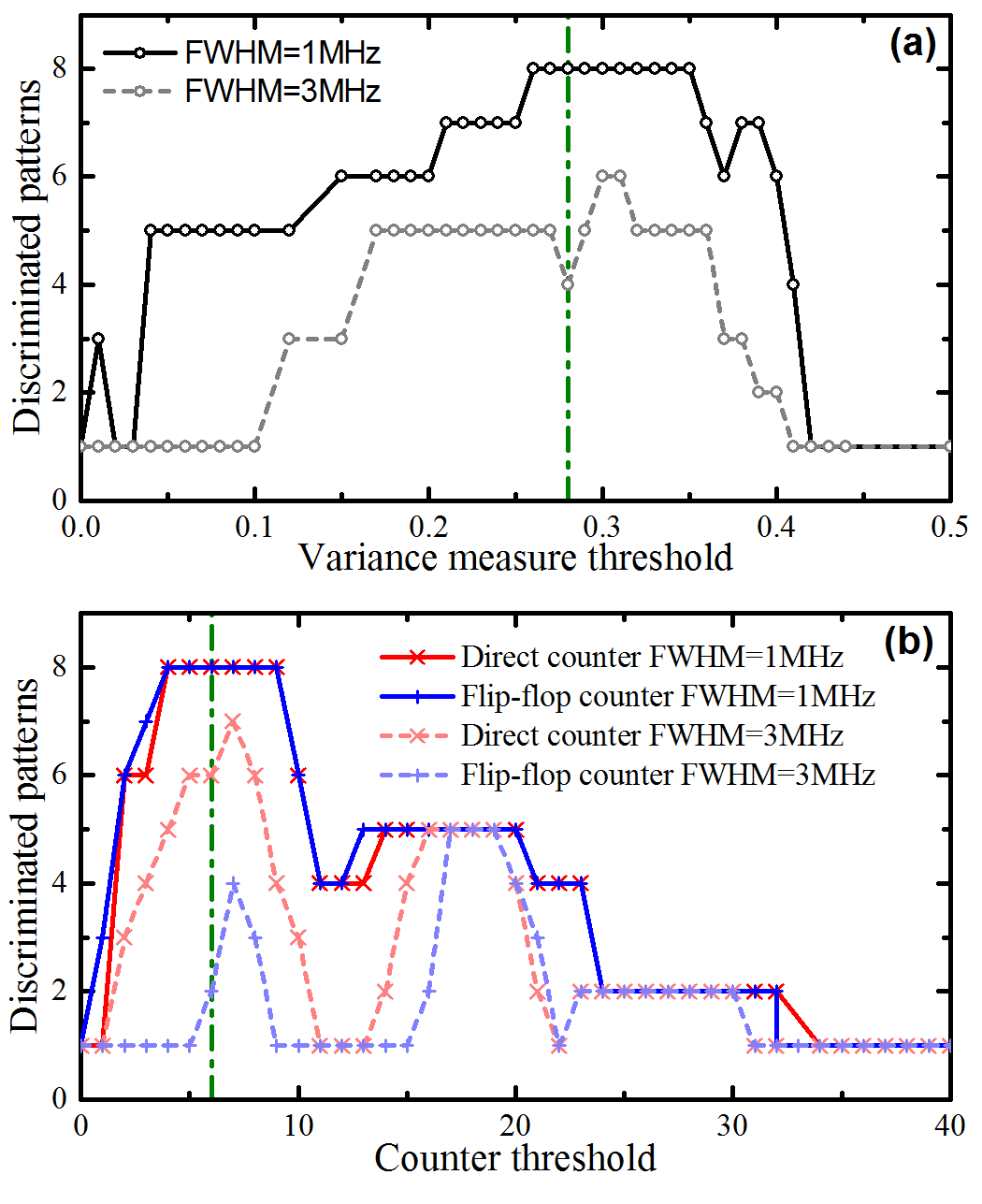}
	\caption{Number of discriminated patterns, (a) as a function of the threshold $\epsilon_{v}$ when using the variance measure technique, (b) as a function of the counter threshold when using the direct counter and flip-flop counter techniques. The FWHM of the oscillators is set to 1\,MHz (solid lines) and 3\,MHz (dotted lines). Vertical green dotted lines indicate the thresholds that were chosen in section~\ref{sec:equivalence} and used in the simulations to obtain readout maps.}
	\label{fig:Patterns_Vs_Thresholds}
\end{figure}

We have seen that in the presence of noise, the three readout methods may suffer difficulties to detect quasi-synchronization. In these conditions, the initial choice of the thresholds, in an oversimplified case and for a noiseless network, should be reconsidered.
We now analyze the impact of the choice of the threshold on the number of recognized patterns.
Fig.~\ref{fig:Patterns_Vs_Thresholds}(a,b) shows the total number of discriminated classes for the $\mathrm{FWHM}=1$\,MHz (solid lines) and for the $\mathrm{FWHM}=3$\,MHz (dotted lines) noisy networks as a function of the variance and counter thresholds.

In the case of low phase noise ($\mathrm{FWHM}=1$\,MHz), all three readout techniques identify the maximum number of patterns (eight) for a reasonable range of thresholds. When increasing the thresholds, the variance based detection method rapidly fails when the threshold $\epsilon_{v}$ is chosen above $0.4$. On the other hand, the thresholds for counter-based methods can be increased even above $20$ without failing (five patterns are still detected), much above the maximum counts observed in ~\ref{sec:equivalence}.

As noticed in section~\ref{sec:ResToNoise}, when the FWHM of oscillators reaches $3$\,MHz, the three readout methods detect different numbers of patterns using the thresholds chosen initially: six for the direct counter scheme, four for the variance-based scheme, and only two for the flip-flop counter scheme. The presented plots show that these initial choices are not adapted to the higher noise case, and that other optimal thresholds can be found.
The direct counter approach still shows the best resilience to noise as up to seven synchronization patterns can be read, while the variance measure approach is limited to detecting up to six patterns. In the counter-based approaches, the optimal thresholds are close to our initial choices, yet the optimal ranges are substantially reduced. For the flip-flop counter scheme, the optimal threshold is found around $18$, confirming that high noise induces the spurious detection of many desynchronization events.

\subsection{Influence of the Evaluation Time}

\begin{figure}[h]
	\centering
	\includegraphics[]{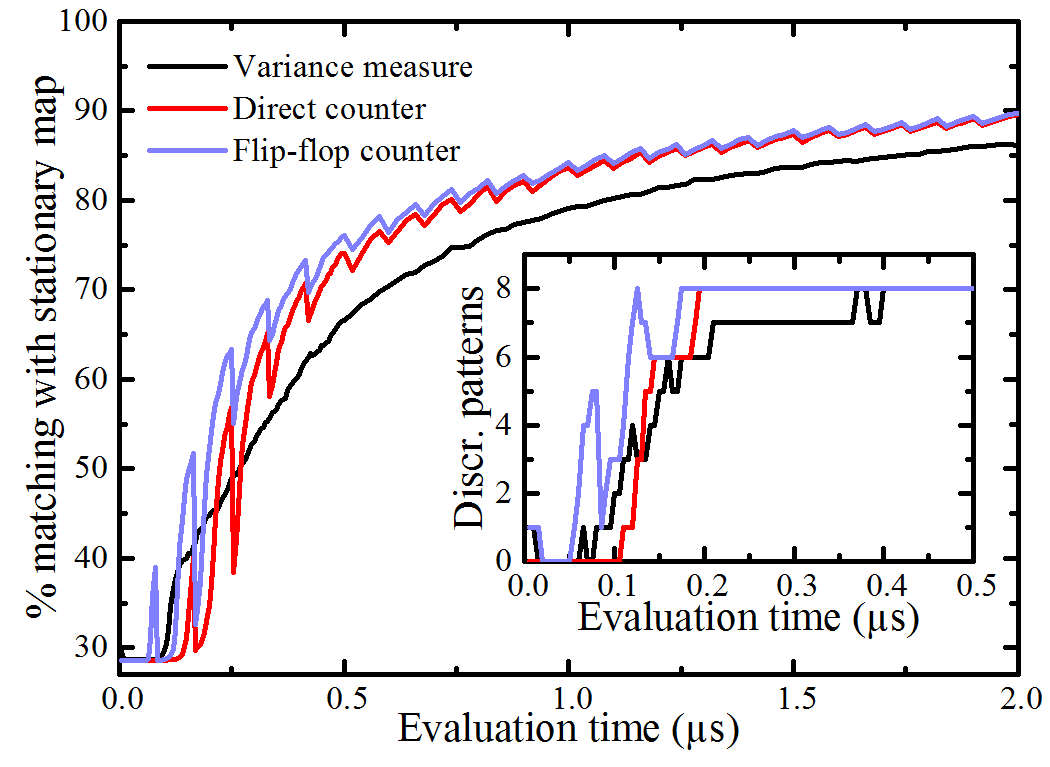}
	\caption{Matching percentage between the readout map obtained for a limited integration time $\tau$ and the readout map obtained for a long $100\mu s$ integration time, as a function of $\tau$ and for the three detection protocols. The inset shows the number of discriminated patterns detected by the three methods as a function of integration time. Simulations are performed with $\mathrm{FWHM}=1$\,MHz.}
	\label{fig:Integration_Time}
\end{figure}

The synchronization evaluation time $\tau$ is an important trade-off for the readout operation, between the speed for recognition and the robustness of the results. 
Simulations of the recognition networks are performed with varying evaluation time, between $0$ and $2\mu$s. The cool-down time is kept to $0.5\mu$s, and the FWHM of the oscillators is $1$\,MHz.
The results are compared to the readout obtained in the case of a long evaluation time $\tau=100\mu s$ for which the readout is considered stationary and further used as a reference.

In Fig.~\ref{fig:Integration_Time}, we plot the percentage of matching points between the obtained readout maps and the reference map, as a function of $\tau$ and for each readout scheme. The figure inset also shows the evolution of the detected number of patterns as a function of $\tau$. For equivalence in the case of the counter-based detection schemes, the thresholds are adjusted as $\tau$ varies so that $\lfloor\epsilon_{\{d,f\}}/\tau\rfloor$ is kept equal to $12\mu$s$^{-1}$.

We observe a fast convergence of the readout maps for evaluation times up to $0.5\mu$s above which the convergence starts slowing down. After $2\mu$s, the counter-based readout schemes reach 90\% matching with the reference, while the variance-based scheme lags behind. For the counter-based methods, the maximum number of patterns (eight) is already detected when $\tau>0.2\mu$s, while $\tau>0.4\mu$s is needed for the variance measure scheme.  In all case, the choice of $\tau=0.5\mu$s then offers a reasonable trade-off as the maximum number of patterns is already discriminated.


\section{Conclusion}

In this work, we have introduced two simple techniques (direct counter and flip-flop counter) for evaluating the output of coupled-oscillator based recognition networks.
Both counters-based approaches are simple to implement in hardware, and were compared to a hardware-implausible, more conventional approach (variance measure).

In situations with oscillators with no or low phase noise, the readouts of all three techniques appear very similar.
Although variance measure is the most complex, it is not the most robust to noise, and may actually identify less synchronization patterns in some situations with intermediate noise levels and very weak coupling between core oscillators. 
The variance measure also appears to converge slower than counter-based approaches, hence requiring longer evaluation time for equivalent precision.


When both counter-based protocols show equivalent results, the flip-flop counter is the best choice for hardware implementation, as it relies simply on an unsigned counter.
However the flip-flop counter protocol appears to fail at high noise levels, when it detects a lot of spurious desycnrhonization events. Nonetheless, the direct counter protocol also shows strong resilience to high noise, again better than the variance measure. The choice between these two techniques should therefore be based on the amount of noise.

These results give credibility to the idea of coupled oscillator-based recognition networks, and open the way for their implementation, either with CMOS technology or emerging nanotechnologies.

\section*{Acknowledgments}
This work is supported by a public grant overseen by the French National Research Agency (ANR) as part of the “Investissement d'Avenir” program (Labex NanoSaclay, reference: ANR-10-LABX-0035), by the ANR MEMOS grant (reference: ANR-14-CE26-0021), and by Ministere  du Developpement durable.
The authors would  like to thank A.~F.~Vincent, A.~Mizrahi and F.~A.~Araujo for fruitful discussions.

\bibliographystyle{IEEEtran}
\bibliography{IEEEabrv,IJCNN_SyncDetection}

\begin{thebibliography}{10}
\providecommand{\url}[1]{#1}
\csname url@samestyle\endcsname
\providecommand{\newblock}{\relax}
\providecommand{\bibinfo}[2]{#2}
\providecommand{\BIBentrySTDinterwordspacing}{\spaceskip=0pt\relax}
\providecommand{\BIBentryALTinterwordstretchfactor}{4}
\providecommand{\BIBentryALTinterwordspacing}{\spaceskip=\fontdimen2\font plus
\BIBentryALTinterwordstretchfactor\fontdimen3\font minus
  \fontdimen4\font\relax}
\providecommand{\BIBforeignlanguage}[2]{{%
\expandafter\ifx\csname l@#1\endcsname\relax
\typeout{** WARNING: IEEEtran.bst: No hyphenation pattern has been}%
\typeout{** loaded for the language `#1'. Using the pattern for}%
\typeout{** the default language instead.}%
\else
\language=\csname l@#1\endcsname
\fi
#2}}
\providecommand{\BIBdecl}{\relax}
\BIBdecl

\bibitem{yen-kuang_chen_convergence_2008}
{Yen-Kuang Chen}, J.~Chhugani, P.~Dubey, C.~Hughes, {Daehyun Kim}, S.~Kumar,
  V.~Lee, A.~Nguyen, and M.~Smelyanskiy, ``Convergence of {Recognition},
  {Mining}, and {Synthesis} {Workloads} and {Its} {Implications},''
  \emph{Proceedings of the IEEE}, vol.~96, no.~5, pp. 790--807, May 2008.

\bibitem{shibata_bio_2009}
T.~Shibata, ``\BIBforeignlanguage{en}{Bio {Inspired} {Architectures} in the
  {Nanoscale} {Integration} {Era}},'' \emph{\BIBforeignlanguage{en}{ECS
  Transactions}}, vol.~25, no.~7, pp. 49--64, Sep. 2009.

\bibitem{indiveri_memory_2015}
G.~Indiveri and S.-C. Liu, ``Memory and information processing in neuromorphic
  systems,'' \emph{Proceedings of the {IEEE}}, vol. 103, no.~8, pp. 1379--1397.

\bibitem{querlioz_bioinspired_2015}
D.~Querlioz, O.~Bichler, A.~Vincent, and C.~Gamrat, ``Bioinspired programming
  of memory devices for implementing an inference engine,'' \emph{Proceedings
  of the {IEEE}}, vol. 103, no.~8, pp. 1398--1416.

\bibitem{merolla_million_2014}
P.~A. Merolla, J.~V. Arthur, R.~Alvarez-Icaza, A.~S. Cassidy, J.~Sawada,
  F.~Akopyan, B.~L. Jackson, N.~Imam, C.~Guo, Y.~Nakamura, B.~Brezzo, I.~Vo,
  S.~K. Esser, R.~Appuswamy, B.~Taba, A.~Amir, M.~D. Flickner, W.~P. Risk,
  R.~Manohar, and D.~S. Modha, ``A million spiking-neuron integrated circuit
  with a scalable communication network and interface,'' \emph{Science}, vol.
  345, no. 6197, pp. 668--673.

\bibitem{benjamin_neurogrid_2014}
B.~Benjamin, P.~Gao, E.~{McQuinn}, S.~Choudhary, A.~Chandrasekaran, J.-M.
  Bussat, R.~Alvarez-Icaza, J.~Arthur, P.~Merolla, and K.~Boahen, ``Neurogrid:
  A mixed-analog-digital multichip system for large-scale neural simulations,''
  \emph{Proceedings of the {IEEE}}, vol. 102, no.~5, pp. 699--716.

\bibitem{roy_brain-inspired_2014}
K.~Roy, M.~Sharad, D.~Fan, and K.~Yogendra, ``Brain-inspired computing with
  spin torque devices,'' in \emph{Design, {Automation} and {Test} in {Europe}
  {Conference} and {Exhibition} ({DATE}), 2014}, Mar. 2014, pp. 1--6.

\bibitem{vincent_synapse_2015}
A.~F. Vincent, J.~Larroque, N.~Locatelli, N.~B. Romdhane, O.~Bichler,
  C.~Gamrat, W.~S. Zhao, J.~O. Klein, S.~Galdin-Retailleau, and D.~Querlioz,
  ``Spin-transfer torque magnetic memory as a stochastic memristive synapse for
  neuromorphic systems,'' \emph{IEEE Transactions on Biomedical Circuits and
  Systems}, vol.~9, no.~2, pp. 166--174, April 2015.

\bibitem{pershin_experimental_2010}
Y.~V. Pershin and M.~Di~Ventra, ``Experimental demonstration of associative
  memory with memristive neural networks,'' \emph{Neural Networks}, vol.~23,
  no.~7, pp. 881--886, Sep. 2010.

\bibitem{prezioso_training_2014}
M.~Prezioso, F.~Merrikh-Bayat, B.~Hoskins, G.~Adam, K.~Likharev, and
  D.~Strukov, ``Training and operation of an integrated neuromorphic network
  based on metal-oxide memristors.'' \emph{Nature}, vol. 521, no. 7550, p.~61,
  2015.

\bibitem{saighi_plasticity_2015}
S.~Sa\"ighi, C.~G. Mayr, T.~Serrano-Gotarredona, H.~Schmidt, G.~Lecerf,
  J.~Tomas, J.~Grollier, S.~Boyn, A.~F. Vincent, D.~Querlioz, S.~La~Barbera,
  F.~Alibart, D.~Vuillaume, O.~Bichler, C.~Gamrat, and B.~Linares-Barranco,
  ``Plasticity in memristive devices for spiking neural networks,''
  \emph{Frontiers in Neuroscience}, vol.~9, Mar. 2015.

\bibitem{bichler_visual_2012}
M.~Bichler, D.~Suri, D.~Querlioz, D.~Vuillaume, B.~De~Salvo, and C.~Gamrat,
  ``{Visual pattern extraction using energy-efficient ''2-PCM synapse''
  neuromorphic architecture},'' \emph{{IEEE Transactions on Electron Devices}},
  vol.~59, pp. 2206--2214, 2012.

\bibitem{axmacher_memory_2006}
N.~Axmacher, F.~Mormann, G.~Fernández, C.~E. Elger, and J.~Fell, ``Memory
  formation by neuronal synchronization,'' \emph{Brain Research Reviews},
  vol.~52, no.~1, pp. 170--182, 2006.

\bibitem{bhowmik_how_2012}
D.~Bhowmik and M.~Shanahan, ``How well do oscillator models capture the
  behaviour of biological neurons?'' in \emph{The 2012 {International} {Joint}
  {Conference} on {Neural} {Networks} ({IJCNN})}, Jun. 2012, pp. 1--8.

\bibitem{hopfield_neural_1982}
J.~J. Hopfield, ``\BIBforeignlanguage{en}{Neural networks and physical systems
  with emergent collective computational abilities},''
  \emph{\BIBforeignlanguage{en}{Proceedings of the National Academy of
  Sciences}}, vol.~79, no.~8, pp. 2554--2558, Apr. 1982.

\bibitem{johannet_specification_1992}
A.~Johannet, L.~Personnaz, G.~Dreyfus, J.-D. Gascuel, and M.~Weinfeld,
  ``Specification and implementation of a digital {Hopfield}-type associative
  memory with on-chip training,'' \emph{IEEE Transactions on Neural Networks},
  vol.~3, no.~4, pp. 529--539, 1992.

\bibitem{hoppensteadt_associative_1997}
F.~Hoppensteadt and E.~Izhikevich, ``Associative memory of weakly connected
  oscillators,'' in \emph{, {International} {Conference} on {Neural}
  {Networks},1997}, vol.~2, 1997, pp. 1135--1138 vol.2.

\bibitem{izhikevich_weakly_1999}
E.~Izhikevich, ``Weakly pulse-coupled oscillators, {FM} interactions,
  synchronization, and oscillatory associative memory,'' \emph{IEEE
  Transactions on Neural Networks}, vol.~10, no.~3, pp. 508--526, 1999.

\bibitem{nikonov_coupled-oscillator_2015}
D.~Nikonov, G.~Csaba, W.~Porod, T.~Shibata, D.~Voils, D.~Hammerstrom, I.~Young,
  and G.~Bourianoff, ``Coupled-{Oscillator} {Associative} {Memory} {Array}
  {Operation} for {Pattern} {Recognition},'' \emph{IEEE Journal on Exploratory
  Solid-State Computational Devices and Circuits}, vol.~1, pp. 85--93, 2015.

\bibitem{levitan_associative_2013}
S.~P. Levitan, Y.~Fang, J.~A. Carpenter, C.~N. Gnegy, N.~S. Janosik,
  S.~Awosika-Olumo, D.~M. Chiarulli, G.~Csaba, and W.~Porod, ``Associative
  processing with coupled oscillators,'' in \emph{Proceedings of the 2013
  International Symposium on Low Power Electronics and Design}, ser. {ISLPED}
  '13.\hskip 1em plus 0.5em minus 0.4em\relax {IEEE} Press, pp. 235--235.

\bibitem{cosp_synchronization_2004}
J.~Cosp, J.~Madrenas, E.~Alarcon, E.~Vidal, and G.~Villar,
  ``\BIBforeignlanguage{en}{Synchronization of {Nonlinear} {Electronic}
  {Oscillators} for {Neural} {Computation}},''
  \emph{\BIBforeignlanguage{en}{IEEE Transactions on Neural Networks}},
  vol.~15, no.~5, pp. 1315--1327, Sep. 2004.

\bibitem{sivilotti_novel_1990}
M.~A. Sivilotti, M.~Emerling, and C.~Mead, ``A {Novel} {Associative} {Memory}
  {Implemented} {Using} {Collective} {Computation},'' in \emph{Artificial
  neural networks: electronic implementations}, N.~Morgan, Ed.\hskip 1em plus
  0.5em minus 0.4em\relax Piscataway, NJ: IEEE, 1990, pp. 11--21.

\bibitem{shibata_cmos_2012}
T.~Shibata, R.~Zhang, S.~Levitan, D.~Nikonov, and G.~Bourianoff, ``{CMOS}
  supporting circuitries for nano-oscillator-based associative memories,'' in
  \emph{2012 13th {International} {Workshop} on {Cellular} {Nanoscale}
  {Networks} and {Their} {Applications} ({CNNA})}, 2012, pp. 1--5.

\bibitem{cotter_computational_2014}
M.~Cotter, Y.~Fang, S.~Levitan, D.~Chiarulli, and V.~Narayanan, ``Computational
  {Architectures} {Based} on {Coupled} {Oscillators},'' in \emph{2014 {IEEE}
  {Computer} {Society} {Annual} {Symposium} on {VLSI} ({ISVLSI})}, Jul. 2014,
  pp. 130--135.

\bibitem{hoppensteadt_synchronization_2000}
F.~C. Hoppensteadt and E.~M. Izhikevich, ``Synchronization of laser
  oscillators, associative memory, and optical neurocomputing,'' \emph{Physical
  Review E}, vol.~62, no.~3, pp. 4010--4013, Sep. 2000.

\bibitem{csaba_spin_2012}
G.~Csaba, M.~Pufall, D.~Nikonov, G.~Bourianoff, A.~Horvath, T.~Roska, and
  W.~Porod, ``Spin torque oscillator models for applications in associative
  memories,'' in \emph{2012 13th {International} {Workshop} on {Cellular}
  {Nanoscale} {Networks} and {Their} {Applications} ({CNNA})}, 2012, pp. 1--2.

\bibitem{fan_injection_2015}
D.~Fan, S.~Maji, K.~Yogendra, M.~Sharad, and K.~Roy, ``Injection-locked spin
  hall-induced coupled-oscillators for energy efficient associative
  computing,'' \emph{Nanotechnology, IEEE Transactions on}, vol.~14, no.~6, pp.
  1083--1093, Nov 2015.

\bibitem{datta_neuro_2014}
S.~Datta, N.~Shukla, M.~Cotter, A.~Parihar, and A.~Raychowdhury, ``Neuro
  inspired computing with coupled relaxation oscillators,'' in \emph{2014 51st
  {ACM}/{EDAC}/{IEEE} {Design} {Automation} {Conference} ({DAC})}, Jun. 2014,
  pp. 1--6.

\bibitem{shukla_pairwise_2014}
N.~Shukla, A.~Parihar, M.~Cotter, M.~Barth, X.~Li, N.~Chandramoorthy, H.~Paik,
  D.~Schlom, V.~Narayanan, A.~Raychowdhury, and S.~Datta, ``Pairwise coupled
  hybrid vanadium dioxide-{MOSFET} ({HVFET}) oscillators for non-boolean
  associative computing,'' in \emph{Electron {Devices} {Meeting} ({IEDM}), 2014
  {IEEE} {International}}, 2014, pp. 28.7.1--28.7.4.

\bibitem{parihar_exploiting_2014}
A.~Parihar, N.~Shukla, S.~Datta, and A.~Raychowdhury, ``Exploiting
  synchronization properties of correlated electron devices in a non-boolean
  computing fabric for template matching,'' \emph{IEEE Journal on Emerging and
  Selected Topics in Circuits and Systems}, vol.~4, no.~4, pp. 450--459, 2014.

\bibitem{sharma_phase_2015}
A.~Sharma, J.~Bain, and J.~Weldon, ``Phase {Coupling} and {Control} of
  {Oxide}-based {Oscillators} for {Neuromorphic} {Computing},'' \emph{IEEE
  Journal on Exploratory Solid-State Computational Devices and Circuits},
  vol.~PP, no.~99, pp. 1--1, 2015.

\bibitem{vassilieva_learning_2011}
E.~Vassilieva, G.~Pinto, J.~Acacio~de Barros, and P.~Suppes, ``Learning
  {Pattern} {Recognition} {Through} {Quasi}-{Synchronization} of {Phase}
  {Oscillators},'' \emph{IEEE Transactions on Neural Networks}, vol.~22, no.~1,
  pp. 84--95, Jan. 2011.

\bibitem{milshtejn_approximate_1975}
G.~Mil’shtejn, ``Approximate integration of stochastic differential
  equations,'' \emph{Theory of Probability \& Its Applications}, vol.~19,
  no.~3, pp. 557--562.

\end{thebibliography}

\end{document}